\documentclass[12pt]{article}
\usepackage{amssymb}
\def\rlx{\relax\leavevmode}
\def\inbar{\vrule height1.5ex width.4pt depth0pt}
\def\IZ{\rlx\hbox{\small \sf Z\kern-.4em Z}}
\def\IR{\rlx\hbox{\rm I\kern-.18em R}}
\def\ID{\rlx\hbox{\rm I\kern-.18em D}}
\def\IC{\rlx\hbox{\,$\inbar\kern-.3em{\rm C}$}}
\def\IN{\rlx\hbox{\rm I\kern-.18em N}}
\def\IP{\rlx\hbox{\rm I\kern-.18em P}}
\def\one{\hbox{{1}\kern-.25em\hbox{l}}}

\def\beq{\begin{equation}}
\def\eeq{\end{equation}}
\def\bea{\begin{eqnarray}}
\def\eea{\end{eqnarray}}
\def\ber{\begin{array}}
\def\eer{\end{array}}

\begin{document}

\begin{titlepage}

December 2000 \hfill{UTAS-PHYS-00-16}\\

\vskip 1.6in
\begin{center}
{\Large {\bf On schizosymmetric superfields and $sl(2/1, {\mathbb C})_{\mathbb R}$ 
supersymmetry  }}
\end{center}

\normalsize
\vskip .4in

\begin{center}
P D Jarvis and K S Fienberg\footnote{
School of Geography and Environmental Studies}, 
\par \vskip .1in \noindent
{\it School of Mathematics and Physics, University of Tasmania}\\
{\it GPO Box 252-21, Hobart Tas 7001, Australia }\\

\end{center}
\par \vskip .3in \noindent

\vspace{1cm}
\noindent
Superfield expansions over four-dimensional graded spacetime 
$(x^{\mu}, \theta^{\nu})$,
with Minkowski coordinates $x$ extended by vector Grassmann 
variables $\theta$, are investigated. By appropriate identification of the physical Lorentz algebra
in the even and odd parts of the superfield,
a typology of `schizofields' containing both integer and 
half-integer spin fields is established. For two of these types, 
identified as 
`gauge potential'-like and `field strength'-like 
schizofields, an $sl(2/1, {\mathbb C})_{\mathbb R}$ supersymmetry at the component 
field level is demonstrated. Prospects for a schizofield calculus, 
and application of these types of fields to the particle spectrum, are 
adumbrated. 

\end{titlepage}

\section{Introduction}
In a previous paper\cite{ModPhysLett} a new paradigm of 
`schizosymmetry' for superfield expansions was proposed. The principle 
was based on the assignment of symmetry algebras (of either spacetime 
or internal supersymmetry) of the type
\begin{equation}
T^{phys} = T^{odd}{\mathbb P}^{o} +  T^{even}{\mathbb P}^{e}
\label{eq:SchizoDef}
\end{equation}
whereby physical generators may be assigned differently on even and odd 
parts of the superfield. In \cite{ModPhysLett} the implications of 
(\ref{eq:SchizoDef}) were explored for spacetime symmetries in 
four dimensions, and candidate superfields carrying both 
integer and half-integer spin component fields were identified.

In the present work the analysis of \cite{ModPhysLett} is extended by 
giving the most general schizofield expansions over graded spacetime in four dimensions,
with coordinates $x^{\mu}$ extended by `vector' Grassmann variables 
$\theta^{\nu}$, with physical spin 
assignments which are compatible with (\ref{eq:SchizoDef}) and 
with spin-statistics (\S 2 below). Further, it is shown in \S 3 that 
each of the three classes identified, types $I$, $I\!I$, $I\!I\!I$ together with their
Grassmann parity-inverted versions $\widetilde{I}$, $\widetilde{I\!I}$, $\widetilde{I\!I\!I}$ 
can be viewed as irreducible 16-dimensional representations of one of the 
super-Lorentz algebras $sl(2/1, {\mathbb C})_{\mathbb R}$ or $osp(3,1/4, 
{\mathbb R})$. For the former, `gauge potential'-like and `field strength'-like 
schizofields are identified, and the generators of $sl(2/1, {\mathbb 
C})_{\mathbb R}$ given explicitly. The prospects for a schizofield calculus, 
and application of these types of fields to the particle spectrum, are 
adumbrated in \S 4 which also includes concluding remarks and outlook. 

Apart from antecedents in general works on supersymmetry and 
superfields (see references in \cite{ModPhysLett}), work related to 
the present approach to superfields is that of \cite{Mankoc}, which 
also develops superfields over vector Grassmann coordinates 
(especially in $d \ge 5$ dimensions) extended 
to local symmetries.
In connection with generalised `graded Lorentz' supersymmetries, the 
paper \cite{Lukierski} should be noted.
Finally, although having the use of vector Grassmann coordinates in 
common, studies of supersymmetric quantum mechanics and the 
index theorem (see for example \cite{Alvarez}) and related 
spinning particle models appear to be different from the present 
schizosymmetric superfields. The present work is partly based on 
\cite{Fienberg}

\section{Typology of schizofields}
In order to introduce the notion of schizosymmetry in the space-time 
context in four dimensions, it is necessary to establish some notation 
on the structure of the Lorentz algebra and its representations. For completeness this is given 
below. This material itself is standard
(see for example \cite{texts}), but
it is applied here specifically to the analysis of the spin content 
of the types of superfields under study.
\subsubsection*{Preliminaries}
Graded spacetime is taken to comprise four dimensional
Minkowski space with standard coordinates $(x^{\mu}) = (x^{0}, x^{1}, 
x^{2}, x^{3})$ and Lorentz metric 
$(\eta_{\mu  \nu}) = \mbox{diag} (+1, -1, -1, -1) $, extended by   
`vector' Grassmann coordinates $(\theta^{\mu}) = (\theta^{0}, \theta^{1}, 
\theta^{2}, \theta^{3})$. General 
superfield expansions are then of the form
\begin{equation}
\Phi(x, \theta) = A(x) + \theta^{\mu} V_{\mu}(x) 
+ {\textstyle \frac 12} \theta^{\mu} \theta^{ \nu} F_{\mu \nu}(x) +  
 {\textstyle \frac 16} \theta^{\mu} \theta^{ \nu} \theta^{\rho} A_{\mu \nu \rho}(x)
+ {\textstyle \frac{1}{24}} \theta^{\mu} \theta^{ \nu} \theta^{\rho} \theta^{ \sigma} 
B_{\mu \nu \rho \sigma}(x)
\end{equation}
with the generators of Lorentz transformations represented 
differentially by\footnote{ 
The usual graded Leibnitz rule
$\frac{\partial}{\partial \theta^{\mu}} \theta^{\nu} = 
{\delta_{\mu}}^{\nu} -  \theta^{\nu} \frac{\partial}{\partial \theta^{\mu}} $
applies.}
$$
L_{\mu \nu} = (x_{\mu}\frac{\partial}{\partial x^{\nu}} - 
x_{\nu}\frac{\partial}{\partial x^{\mu}}) + 
(\theta_{\mu}\frac{\partial}{\partial \theta^{\nu}} - 
\theta_{\nu}\frac{\partial}{\partial \theta^{\mu}}). 
$$
That such superfields may also carry spinors is 
evident from the fact that representations of the
Dirac algebra are available,
\begin{eqnarray}
\gamma^{\pm}_{\mu} &=& (\theta_{\mu} \pm \frac{\partial}{\partial 
\theta^{\mu}}), \nonumber \\
\{ \gamma^{\pm}_{\mu}, \gamma^{\pm}_{\nu} \}_{+} &=& \pm 2 \eta_{\mu 
\nu}, \nonumber \\
\{ \gamma^{+}_{\mu}, \gamma^{-}_{\nu} \}_{+} &=& 0.
\label{eq:DiracAlgs}
 \end{eqnarray}
with, for example $\gamma_{5}^{\pm} = i 
\gamma_{0}^{\pm} \gamma_{1}^{\pm}\gamma_{2}^{\pm}\gamma_{3}^{\pm}$ 
given by
\begin{equation}
\gamma_{5}^{\pm}          
= {\textstyle \frac{i}{24}} \epsilon^{\mu \nu \rho \sigma}(
\theta_{\mu}\theta_{\nu}\theta_{\rho}\theta_{\sigma} 
\pm 4 \theta_{\mu}\theta_{\nu} \theta_{\rho} \partial_{\sigma}
+ 6 \theta_{\mu}\theta_{\nu}\partial_{\rho}\partial_{\sigma} 
\pm 4 \theta_{\mu} \partial_{\nu} \partial_{\rho}\partial_{\sigma}
+ \partial_{\mu}\partial_{\nu} \partial_{\rho}\partial_{\sigma}).
\end{equation}
The two (anti-commuting, \textit{graded}) representations of the Dirac algebra  
lead to two inequivalent assignments of the spin part of the Lorentz algebra
\begin{eqnarray*}
L^{\pm}_{\mu \nu} &=& \pm \textstyle{\frac 14} [ \gamma^{\pm}_{\mu}, 
\gamma^{\pm}_{\nu} ]_{-}, \nonumber \\
L^{\pm}_{\mu \nu} &=& \textstyle{\frac 12} (\theta_{\mu}
{\displaystyle \frac{\partial}{\partial \theta^{\nu}}} - 
\theta_{\nu}{\displaystyle \frac{\partial}{\partial \theta^{\mu}}}) \pm 
\textstyle{\frac 12} (\theta_{\mu}\theta_{\nu} + 
{\displaystyle \frac{\partial}{\partial \theta^{\mu}}\frac{\partial}{\partial \theta^{\nu}}} )
\nonumber \\
{[} L_{\mu \nu}, L_{\rho \sigma} {]}_{-} &=& \eta_{\rho \nu} L_{\mu \sigma}
- \eta_{\rho \mu} L_{\nu \sigma} - \eta_{\sigma \nu} L_{\mu \rho}
+ \eta_{\sigma \mu} L_{\nu \rho}
\end{eqnarray*}
In the usual way, one can effect the decomposition of the
$L^{\pm}_{\mu \nu}$
into self-dual and anti-self dual combinations,
$$
L^{L/R}_{\mu \nu} = \frac 12 (  L_{\mu \nu} 
\pm \frac{i}{2}{\epsilon_{\mu \nu}}^{\rho \sigma}L_{\rho \sigma})
\equiv \frac 12 (  L_{\mu \nu} \pm i {\tilde{L}}_{\rho \sigma})
$$
which satisfy the same commutation relations as $ L_{\mu \nu}$ but mutually 
commute.
In order to see the implications of these definitions it is convenient 
also to introduce the Weyl representation of the Dirac 
algebra. Dirac spinors are  represented as
$\psi = \mbox{}^{T}\!\left(  u_{a}, \; {\bar{v}}^{\dot{a}} \right)$
with 
\begin{eqnarray*}
\! \gamma_{\mu} = \left( \begin{array}{cc} 0 & {\bar{\sigma}}_{\mu} \\
                                   \sigma_{\mu} & 0 \end{array} 
                                   \right), &\quad &                                    
\textstyle{\frac 14} [ \gamma_{\mu}, \gamma_{\nu} ]  =  
\left( \begin{array}{cc} \sigma_{\mu \nu} & 0 \\
         0 & {\bar{\sigma}}_{\mu \nu} \end{array} \right), \nonumber \\
{ \sigma_{\mu \nu  a}}^{b} \equiv \textstyle{\frac 14} {{({\bar{\sigma}}_{\mu}\sigma_{\nu}
                      \! - \! {\bar{\sigma}}_{\nu}\sigma_{\mu} 
                      )}_{a}}^{b} & \quad &    
 {\bar{\sigma}}_{\mu \nu \dot{b}}^{\dot{a}} \equiv \textstyle{\frac 14} 
 {{({{\sigma}}_{\mu}\bar{\sigma}_{\nu}
                      \! - \! {{\sigma}}_{\nu}\bar{\sigma}_{\mu} )}^{\dot{a}}}_{\dot{b}}
                      \end{eqnarray*}
Then the self-dual parts are projected simply via
\begin{eqnarray*}
J_{a b} \equiv & \textstyle{\frac 14} {(\sigma^{\mu \nu}L_{\mu \nu})}_{a b} &
\equiv \textstyle{\frac 14} {(\sigma^{\mu \nu}L^{L}_{\mu \nu})}_{a b} \\
\bar{J}_{\dot{a} \dot{b}} \equiv & 
\frac 14 {(\bar{\sigma}^{\mu \nu}L_{\mu \nu})}_{\dot{a} \dot{b}}&
\equiv \textstyle{\frac 14} {(\bar{\sigma}^{\mu \nu}L^{R}_{\mu \nu})}_{\dot{a} \dot{b}}
\end{eqnarray*}
because ${\tilde{\sigma}}_{\mu \nu} = -i \sigma_{\mu \nu}$
and ${\tilde{\bar{\sigma}}}_{\mu \nu} = i {\bar{\sigma}}_{\mu \nu}$.
The identification of $so(3,1)$
with $ sl(2, {\mathbb C})_{\mathbb R}$ is completed by 
examining the structure of $sp(2) \sim sl(2)$ generated by
$\{ K_{ab} = K_{ba}; a,b =1, 2 \}$,
$$
[K_{ab}, K_{cd}] _{-} = \varepsilon_{cb}K_{ad} + \varepsilon_{ca}K_{bd}
                      +  \varepsilon_{da}K_{bc} +  
                      \varepsilon_{db}K_{ac},
$$
where $\varepsilon_{12} = 1 = - \varepsilon_{21}$ and 
$\varepsilon_{11} = 0 = \varepsilon_{22}$. 
Then $sp(2, {\mathbb C})_{\mathbb R}$ is spanned by  $\{ K_{ab},\; 
K'_{ab}\equiv iK_{ab} \}$
as a real Lie algebra. In its complexification, we can define
$$
J_{ab} := \textstyle{\frac 12} (K_{ab}+ iK'_{ab}), 
\quad J_{\dot{a} \dot{b}} := \textstyle{\frac 12} (K_{\dot{a} \dot{b}}- iK'_{\dot{a} \dot{b}})
$$
which mutually commute and satisfy the same algebra as $\{ K_{ab} \}$.
In practice, as is well known, it is convenient to label irreducible 
representations\footnote{For example, for a 
Dirac spinor $\sim (\frac 12, 0) + (0,\frac 12)$, 
4-vector $\sim (\frac 12, \frac 12)$, 
antisymmetric tensor $\sim (1, 0) + (0,1)$ and so on.} of the Lorentz algebra via the complexification
$SO(3,1)_{\mathbb C} \simeq sp(2, {\mathbb C}) \simeq sl(2)_{L} \oplus 
		sl(2)_{R}$. 

Returning to the superfield expansion of $\Phi(x, \theta)$, note in the Dirac algebra 
$$
\gamma_{5} \cdot \textstyle{\frac 14} [\gamma_{\mu},\gamma_{\nu}] = 
\frac{i}{2}{\epsilon_{\mu\nu}}^{\rho \sigma} 
\textstyle{\frac 14} [\gamma_{\rho},\gamma_{\sigma}]
$$
and also
\begin{eqnarray*}
\textstyle{\frac 12}(1\!+\!\gamma_{5}) \cdot \left( \begin{array}{cc} \sigma_{\mu \nu} & 0 \\
         0 & {\bar{\sigma}}_{\mu \nu} \end{array} \right) &=&
         \left( \begin{array}{cc} \sigma_{\mu \nu} & 0 \\
         0 & 0 \end{array} \right), \\
\textstyle{\frac 12}(1\!-\!\gamma_{5}) \cdot \left( \begin{array}{cc} \sigma_{\mu \nu} & 0 \\
         0 & {\bar{\sigma}}_{\mu \nu} \end{array} \right) &=&
         \left( \begin{array}{cc} 0 & 0 \\
         0 & {\bar{\sigma}}_{\mu \nu} \end{array} \right)
\end{eqnarray*}
so that the $\frac 12(1 \!\pm \! \gamma^{\pm}_{5})$ projections can be 
used to decompose 
superfield components in $\Phi(x, \theta)$ with respect to 
$SO(3,1)^{+}+SO(3,1)^{-} \simeq sl(2)_{L}^{+}+ 
sl(2)_{R}^{+}+sl(2)_{L}^{-}+ sl(2)_{R}^{-}$ and determine the 
complete spectrum 
of $(j_{1},j_{2})^{\pm}$. The results are shown in table 
\ref{tbl:ThetaMonomials}\footnote{
Where $\theta^{\mu \nu} = {\textstyle \frac 12 }\theta^{\mu}\theta^{\nu}$,
$\tilde{\theta}^{\mu\nu} = {\textstyle \frac 12} \varepsilon^{\mu\nu\rho\sigma}
\theta_{\rho}\theta_{\sigma}$,
$\tilde{\theta}^{\mu} = {\textstyle \frac 16} \varepsilon^{\mu\nu\rho\sigma}
\theta_{\nu}\theta_{\rho}\theta_{\sigma}$,
$\theta^{4} = {\textstyle \frac {1}{24}} \varepsilon^{\mu\nu\rho\sigma}
\theta_{\mu}\theta_{\nu}\theta_{\rho}\theta_{\sigma}$
}.
\begin{table}[tbp]
	\centering
	\caption{Schizofield spherical harmonics: $\theta$-polynomials covariant with respect to 
	$SO(3,1)^{+}+SO(3,1)^{-} \simeq sl(2)_{L}^{+}+ 
sl(2)_{R}^{+}+sl(2)_{L}^{-}+ sl(2)_{R}^{-}$ } \vspace*{.3cm} 
	\begin{tabular}{|c||c|c|}
         	\hline
         	$\theta$-polynomial & $\gamma_{5}^{+} \; \gamma_{5}^{-}$ & 
         	$so(3,1)^{+}\oplus so(3,1)^{-}$ \\
         	\hline
         	\hline
         	$\frac 12(1+i\theta^{4})$ & $+ \; +$ & $(0, \frac 12) \otimes (0, \frac 12)$ \\
         	\hline
         	$\frac 12(1-\theta^{4})$ & $- \; -$ & $( \frac 12, 0) \otimes (\frac 12, 0)$ \\
         	\hline
         	$\frac 12(\theta^{\mu}+i\tilde{\theta}^{\mu})$ & $- \; +$ & 
         	$( \frac 12, 0) \otimes (0, \frac 12)$ \\
         	\hline
         	$\frac 12(\theta^{\mu}-i\tilde{\theta}^{\mu})$ & $+ \; -$ & 
         	$(0, \frac 12) \otimes ( \frac 12, 0)$ \\
         	\hline
         	$\frac 12(\theta^{\mu \nu}+i\tilde{\theta}^{\mu \nu})$ & $- 
         	\; -$ & $( \frac 12, 0) \otimes ( \frac 12, 0)$ \\
         	\hline
         	$\frac 12(\theta^{\mu \nu}-i\tilde{\theta}^{\mu \nu})$ & $+ 
         	\; +$ & $(0, \frac 12) \otimes (0, \frac 12)$ \\
         	\hline
         \end{tabular}
	\label{tbl:ThetaMonomials}
\end{table}
         
\subsubsection*{Schizosymmetry - a new paradigm 
for superfield expansions }
The principle of schizosymmetry is now implemented by the choice of 
physical Lorentz algebra in extended spacetime,
\begin{equation}
	L^{phys} = L^{odd}{\mathbb P}^{o} +   L^{even}{\mathbb P}^{e}
	\label{eq:Lorentzschizo}
\end{equation}
with projection operators defined through the number operator
${\cal N} \equiv \theta^{\mu} \partial_{\mu}$, for example
\begin{eqnarray}
{\mathbb P}^{e} &=& {\textstyle \frac 13 }({\cal N} -1)({\cal N} -3)({\cal N}^{2} - 4{\cal 
N}+1) \nonumber \\
 &=&  {\textstyle \frac 13} (\theta^{\mu}\theta^{\nu}\theta^{\rho}\theta^{\sigma}
   \partial_{\sigma}\partial_{\rho}\partial_{\nu}\partial_{\mu}
 -2\theta^{\mu}\theta^{\nu}\theta^{\rho}\partial_{\rho}\partial_{\nu}\partial_{\mu} 
  +3 \theta^{\mu}\theta^{\nu}\partial_{\nu}\partial_{\mu} -3 \theta^{\mu} 
 \partial_{\mu}+3); \nonumber  \\
{\mathbb P}^{o} &=& -{\textstyle \frac 13 }{\cal N}({\cal N} -2)^{2}({\cal N} -4) \nonumber \\
   &=& -{\textstyle \frac 13 }(\theta^{\mu}\theta^{\nu}\theta^{\rho}\theta^{\sigma}
   \partial_{\sigma}\partial_{\rho}\partial_{\nu}\partial_{\mu}
 -2\theta^{\mu}\theta^{\nu}\theta^{\rho}\partial_{\rho}\partial_{\nu}\partial_{\mu} 
  +3 \theta^{\mu}\theta^{\nu}\partial_{\nu}\partial_{\mu} -3 \theta^{\mu} 
 \partial_{\mu}); \nonumber \\
{\mathbb P}^{(0)} &=& {\textstyle \frac{1}{24}} ({\cal N} -1)({\cal N} -2)({\cal N} 
-3)({\cal N} -4)\nonumber  \\
   &\equiv& \partial_{\sigma}\partial_{\rho}\partial_{\nu}\partial_{\mu}
   \theta^{\mu}\theta^{\nu}\theta^{\rho}\theta^{\sigma};
   \label{eq:projectors}
\end{eqnarray}
the projection ${\mathbb P}^{(0)}$ on to the $\theta$-independent term of the 
superfield will be required in \S 3 below.
In fact, all acceptable embeddings of $L^{phys}_{\mu\nu}$ correspond to consistent reductions 
of the eight-dimensional even and odd projections of the schizofield into tensor and spinor 
representations respectively. This counting problem turns out only to 
have essentially three solutions, shown in table 
\ref{tbl:schizoclasses} as classes $I$, $I\!I$, $I\!I\!I$.
\begin{table}[tbp]
	\centering
	\caption{Component field content of `even' type 
	schizofields of classes $I$, $I\!I$, $I\!I\!I$ 
	(for the `odd' types, the component field contents of the even and 
	odd parts of the superfields are reversed).}
	\begin{tabular}{|c||c|c|}
		\hline
		Class & even & odd  \\
		\hline
		\hline
		 $I$ & $(1,0)\!+\!(0,1)\!+\!2(0,0)$  & 
		 $2[(\frac 12 ,0)\!+\!(0,\frac 12)]$ \\
		\hline
		$I\!I$  & $( \frac 12 ,\frac 12 )\!+\!4(0,0)$  
		&  $2[(\frac 12 ,0)\!+\!(0,\frac 12)]$  \\
		\hline
		$I\!I\!I$ & $2( \frac 12,\frac 12 ) $ &
		  $2[(\frac 12 ,0)\!+\!(0,\frac 12)]$  \\
		\hline
	\end{tabular}
	\label{tbl:schizoclasses}
\end{table}
It is convenient also to introduce the corresponding Grassmann odd schizofield
classes $\widetilde{I}$, $\widetilde{I\!I}$, $\widetilde{I\!I\!I}$ corresponding to the same 
representation content, but with grading reversed. Note that only $so(3,1)$ irreps $(j,j)$ or 
$(j_{1},j_{2})+(j_{2},j_{1})$ appear, and not
cases such as $3(\frac 12 ,0) + (0,\frac 12)$ or $(\frac 12 ,0) + 2(0,0)$. 
Each of classes $I$, $I\!I$, $I\!I\!I$ corresponds to a specific assignment of 
Lorentz algebra of the type (\ref{eq:Lorentzschizo}), for example 
\begin{equation}
	L_{\mu\nu}^{I}= L_{\mu\nu}^{diag}{\mathbb P}^{e}+  L_{\mu\nu}^{+}{\mathbb 
	P}^{ o}
	\label{eq:schizoIdefn}
\end{equation}
with similar representative identifications for the other classes. 
Some of these exploit the semisimple nature of the Lorentz algebra in 
four dimensions, in that the schizosymmetric identifications of type 
(\ref{eq:Lorentzschizo}) are different for $sl(2)^{L}$ and $sl(2)^{R}$.
The explicit physical field content is manifested by decoupling the 
$\mbox{}^{+}$ and $\mbox{}^{-}$ covariance via van der Waerden notation
(compare (\ref{tbl:ThetaMonomials})), yielding
\begin{eqnarray}
\Phi_{I} &=& A +  \theta^{4} B
+ \theta^{\mu \nu}F_{\mu \nu} + 
\textstyle{\frac 12}(\theta^{\mu}+i\tilde{\theta}^{\mu})
\sigma_{\mu}^{a \dot{\alpha}}u_{a \dot{\alpha}} +  \nonumber \\
&& \quad \quad \textstyle{\frac 12} (\theta^{\mu}-i\tilde{\theta}^{\mu})
{\bar{\sigma}}^{\mu}_{\alpha \dot{a}}{\bar{v}}^{\dot{a}\alpha},  \nonumber \\
\Phi_{I\!I} &=& \textstyle{\frac 14} (1-i\theta^{4})\varepsilon^{a 
\alpha}\sigma^{\mu}_{a \dot{\alpha}}V_{\mu} + 
\textstyle{\frac 14} (1+i\theta^{4})\varepsilon^{\dot{a}
\dot{\alpha}}S_{\dot{a}\dot{\alpha}} \nonumber  \\
&& + \textstyle{\frac 12} (\theta^{\mu \nu}-i\tilde{\theta}^{\mu \nu})
\sigma^{a {\alpha}}_{\mu\nu}\sigma^{\lambda}_{a \dot{\alpha}}V_{\lambda} 
 + \textstyle{\frac 12} (\theta^{\mu \nu}+i\tilde{\theta}^{\mu \nu})
\bar{\sigma}^{\dot{a} \dot{\alpha}}_{\mu\nu} S_{\dot{a}\dot{\alpha}} \nonumber  \\
&&  + \textstyle{\frac 12}(\theta^{\mu}+i\tilde{\theta}^{\mu})
\sigma_{\mu}^{a \dot{\alpha}}u_{a\dot{\alpha}} + 
\textstyle{\frac 12} (\theta^{\mu}-i\tilde{\theta}^{\mu})
{\bar{\sigma}}^{\mu}_{\alpha \dot{a}}{\bar{v}}^{\dot{a} \alpha}, \nonumber  \\
\widetilde{\Phi}_{I\!I\!I} &=&
\theta^{\mu}V_{\mu} + \tilde{\theta}^{\mu}A_{\mu}  \nonumber \\
&& + \textstyle{\frac 12} (\theta^{\mu \nu}+i\tilde{\theta}^{\mu \nu}) \sigma^{a 
\alpha}_{\mu\nu}u_{a \alpha}
+ \textstyle{\frac 12} (\theta_{\mu \nu}-i\tilde{\theta}_{\mu \nu}) 
\bar{\sigma}_{\dot{a} \dot{\alpha}}^{\mu\nu}\bar{v}^{\dot{a} \dot{\alpha}}. 
\label{eq:PhiExpansions}
\end{eqnarray}

\section{Super-Lorentz symmetry $sl(2/1,{\mathbb C})_{\mathbb R}$ }
\subsubsection*{Supersymmetry?}

Thus far the admissible schizosymmetric superfields have been derived from 
considerations of Lorentz covariance and consistency with 
spin-statistics, with their even and odd parts considered separately.
It is natural however to look for 
superalgebras for which $\Phi(x,\theta)$ is an irreducible 
representation, and hence invoke a supersymmetric unification at least 
at the component field level. One such candidate is by default
the superalgebra $gl(8/8)$ 
generated by all operators $p(\theta) q(\partial)$ for some polynomials $p,q$
in $\theta^{\mu}$ and derivatives, which certainly has a 
16-dimensional defining representation. However, $gl(8/8)$ is likely 
to be too big to be a useful kinematical superalgebra in further 
constructions such as a schizofield calculus.
More reasonably one can ask for the \textit{smallest} superalgebra
containing the Lorentz algebra for which
for which $\Phi(x,\theta)$ is an irreducible 
representation.

\subsubsection*{Schizofields as representations of $sl(2/1,{\mathbb C})_{\mathbb R}$} 
Consider in this context $sl(2/1, {\mathbb C})_{\mathbb R}$,
a natural supersymmetric grading of the Lorentz algebra. From the 
well known $sl(2/1) \supset sl(2) + U(1)$ representations 
(written $j_{z}$, with charge quantum number $z$ as subscript to spin 
content):
\begin{eqnarray*}
{\mathbf 3} &\downarrow& {\textstyle \frac{1}{2}}_{\frac 12} + 0_{1} \\
{\mathbf 4} &\downarrow& {\textstyle \frac{1}{2}}_{z} + 0_{z+\frac{1}{2}} + 0_{z-\frac {1}{2}} \\
{\mathbf 8} &\downarrow& {1}_{0} + 0_{0} + 
 {\textstyle {\frac {1}{2}} }_{1}+{\textstyle {\frac {1}{2}}}_{-1} 
\end{eqnarray*}
one infers\footnote{As usual, $sl(2/1, {\mathbb C})  \sim sl(2/1)^{L} + 
sl(2/1)^{R}$.} the following representations of $sl(2/1, {\mathbb C})_{\mathbb R}$
(choosing $z=0$):
\begin{eqnarray}
I \;\;  ({\mathbf 8},0)\! + \!(0, {\mathbf 8}) & \downarrow &
2(0,0)_{0,0}\!+\!  (1,0)_{0,0}\!+\!  (0,1)_{0,0} + \nonumber \\
&&[({\textstyle {\frac {1}{2}}},0)_{\pm 1, 0}\!+\! (0,{\textstyle {\frac 
{1}{2}}})_{0,\pm 1 }] \nonumber \\
I\!I \quad \quad \quad  ({\mathbf 4},{\mathbf 4})& \downarrow & 
({\textstyle {\frac {1}{2}}},{\textstyle {\frac {1}{2}}})_{0,0} \!+\! 
(0,0)_{\pm \frac {1}{2}, \pm \frac {1}{2}}\!+\! (0,0)_{\pm \frac {1}{2}, \mp \frac 
{1}{2}}\!+\!   \nonumber \\
&& [({\textstyle {\frac {1}{2}}},0)_{\pm \frac {1}{2}, 0}\!+\! 
(0,{\textstyle {\frac {1}{2}}})_{0,\pm \frac {1}{2} } ] 
\label{eq:sl2/1Cirreps}
\end{eqnarray}
which precisely coincide with the component field content of the 
corresponding schizofield classes\footnote{
The type $I\!I\!I$ schizofields are associated with an embedding of 
the Lorentz algebra in the orthosymplectic superalgebra 
$osp(3,1/4, {\mathbb R})$.}.

Explicitly, consider $sl(2/1) \simeq osp(2/2)$ spanned by $ 
\{ K_{ab},\; Q_{a \pm},\; Y,\; a, b =1,2 \}$:
\begin{eqnarray*}
{[} K_{ab}, K_{cd}{]}_{-} &=& \varepsilon_{cb}K_{ad} + \varepsilon_{ca}K_{bd}
                      +  \varepsilon_{da}K_{bc} +  \varepsilon_{db}K_{ac} \\
{[} K_{ab}, Q_{c \pm}{]}_{-} &=& \varepsilon_{cb}Q_{a \pm} + 
\varepsilon_{ca}Q_{b \pm} \\
{[}Y, Q_{a \pm}{]}_{-} &=& \pm {\textstyle \frac 12} Q_{a \pm} \\
\{ Q_{a \pm}, Q_{b \pm} \}_{+} &=& - K_{ab} + 2 \varepsilon_{ab}Y 
\end{eqnarray*}
The super-Lorentz algebra $sl(2/1,{\mathbb C})_{\mathbb R}$ is spanned by 
combinations such as $\frac 12 (K \pm i K')$ and so on, giving commuting left and right 
parts $J_{ab},\; S_{a \pm},\; Z^{L};\; \bar{J}_{\dot{a} \dot{b}},\;
\bar{S}_{\dot{a}}, \; Z^{R}$ with the same algebra as above. 
Explicit matrix elements for representations of 
$sl(2/1)$ can be derived by acting  
the above generators on an abstract 
basis. In particular,   
the adjoint representation ${\mathbf 8}$ follows from the structure constants.
With basis $\{ \Theta_{ab},\; \Theta,\; \vartheta_{a \pm} \}$ the 
action is:
\begin{eqnarray}
K_{ab}\cdot \Theta_{cd} & = &  \varepsilon_{cb}\Theta_{ad} + \varepsilon_{ca}\Theta_{bd}
                      +  \varepsilon_{da}\Theta_{bc} +  
                      \varepsilon_{db}\Theta_{ac}  \nonumber \\
K_{ab} \cdot \vartheta_{c \pm}  &=&  \varepsilon_{cb}\vartheta_{a \pm} + 
\varepsilon_{ca}\vartheta_{b \pm} \nonumber  \\
Y \cdot  \vartheta_{a \pm} &=& \pm {\textstyle \frac 12} Q_{a \pm}  \nonumber  \\
Q_{a \pm}\cdot \Theta_{bc} & = &  \varepsilon_{ba}\vartheta_{c \pm} 
+  \varepsilon_{ca}\vartheta_{b \pm} \nonumber \\
Q_{a \pm}\cdot \Theta & = & \mp {\textstyle \frac 12}\vartheta_{a \pm} \nonumber  \\
Q_{a \pm}\cdot \vartheta_{b \pm} & = & 0 \nonumber  \\
Q_{a \pm}\cdot \vartheta_{b \mp} & = &- \Theta_{ab}+ \varepsilon_{ab} \Theta.
\label{eq:8ModuleAction}  
\end{eqnarray}
In order to establish how the supersymmetry acts on type $I$ and $I\!I$ 
schizofields, and thus explicitly establish    
 $\Phi_{I}$ and $\Phi_{I\!I}$ as $({\mathbf 8},0) \! + \! (0, {\mathbf 
 8})$ and $({\mathbf 4},{\mathbf 4})$ representations respectively,
a concrete realisation of such modules must be given 
at the superfield level.
The previously identified $\theta$-polynomials (table 
\ref{tbl:ThetaMonomials}) are equivalent to the abstract basis vectors,
and appropriately defined operators $p(\theta) q(\partial)$
having the appropriate matrix elements establish the embedding of 
$sl(2/1)$ in $gl(8/8)$. 

The procedure is illustrated for the $({\mathbf 8},0) \! + \! (0, {\mathbf 8})$,
with the $({\mathbf 4},{\mathbf 4})$ left as a similar calculation. 
Firstly, the identification of the component field content of 
$\Phi_{I}$ (see (\ref{eq:PhiExpansions}))
must be completed by assigning two additional additive quantum numbers consistently 
with (\ref{eq:sl2/1Cirreps}). However, from the right-hand column of 
table \ref{tbl:ThetaMonomials}, it can be seen that the correct
choice is to take these as magnetic
quantum numbers from the {\it commuting} $sl(2)$ factors in each 
sector ($\mbox{}^{L}$ or $\mbox{}^{R}$, respectively), so that the 
additional abelian generators are
\begin{eqnarray} 
Z^{L} &=& {\bar{J}}^{-}_{\dot{1}\dot{2}}\;{\mathbb P}^{o}, \nonumber \\
Z^{R} &=&  {J}^{-}_{ 1 2}\; {\mathbb P}^{o}.
\end{eqnarray}
For a basis set corresponding to (\ref{eq:8ModuleAction}), define
the following polynomials in $\theta^{\mu}$
(see table \ref{tbl:ThetaMonomials} and associated text) 
\begin{eqnarray}
\Theta &= & \textstyle{\frac 12}(1-i\theta^{4}) \nonumber \\
\Theta_{ab} &= &  \textstyle{\frac 12}(\theta^{\mu \nu}+i\tilde{\theta}^{\mu \nu}) 
(\sigma_{\mu\nu})_{ab} \nonumber \\
\vartheta_{a +} &= &  \textstyle{\frac 12}(\theta^{\mu}+i\tilde{\theta}^{\mu}) 
\sigma^{\mu}_{\dot{1} a} \nonumber \\
\vartheta_{a -} &= & \textstyle{\frac 12}(\theta^{\mu}+i\tilde{\theta}^{\mu}) 
\sigma^{\mu}_{\dot{2} a}
\label{eq:8ThetaBasis}
\end{eqnarray}
together with the corresponding conjugates $\bar{\Theta}_{\dot{a}\dot{b}}, 
\bar{\Theta}, \bar{\vartheta}_{\dot{a} {\alpha}}$. Further introduce the 
corresponding polynomials in $\partial^{\theta}_{\mu}$, namely
$\Delta_{ab}, \Delta, \varrho_{a \dot{\alpha}}$, say, and their 
conjugates:
\begin{eqnarray}
\Delta &= & \textstyle{\frac 12}(1-i\partial^{4}) \nonumber \\
\Delta_{ab} &= &  \textstyle{\frac 12}(\partial^{\mu \nu}+i\tilde{\partial}^{\mu \nu}) 
(\sigma_{\mu\nu})_{ab} \nonumber \\
\varrho_{a +} &= &  \textstyle{\frac 12}(\partial^{\mu}+i\tilde{\partial}^{\mu}) 
\sigma^{\mu}_{\dot{1} a} \nonumber \\
\varrho_{a -} &= & \textstyle{\frac 12}(\partial^{\mu}+i\tilde{\partial}^{\mu}) \sigma^{\mu}_{\dot{2} a}
\label{eq:8DiffThetaBasis}
\end{eqnarray}
As has been pointed out by Eyal\cite{Eyal} (see also 
\cite{DelbourgoJarvisWarner}), in the superfield context, 
a basis of $gl(8/8)$ corresponding to elementary matrices must be 
introduced with the use of appropriate zero projector ${\mathbb P}^{(0)}$ (see 
(\ref{eq:projectors}) above). For
polynomials $q(\theta)$, $p(\theta)$, set 
$$
E_{pq} = p(\theta){\mathbb P}^{(0)}q(\partial)
$$
which maps between $q$ and $p$ ({\it with all other matrix elements zero}). 
Hence the schizosymmetric assignment of physical Lorentz generators on class $I$ superfields
is completed by the following generators of the super-Lorentz algebra 
establishing the embedding of the latter in $gl(8/8)$, 
\begin{eqnarray*}
Q_{a \pm}&=& (- \Theta_{ab} + 2 \varepsilon_{ab} \Theta) {\mathbb P}^{(0)} 
\varrho^{b \mp} +
\vartheta^{b \pm}{\mathbb P}^{(0)}(2 \varepsilon_{ab}\Delta - \Delta_{ab}); \nonumber \\
\overline{Q}_{\dot{a} \pm}&= &
(- \overline{\Theta}_{\dot{a}\dot{b}} + 2 \varepsilon_{\dot{a}\dot{b}} \overline{\Theta}) {\mathbb P}^{(0)} 
\overline{\varrho}^{\dot{b} \mp} +
\overline{\vartheta}^{\dot{b} \pm}{\mathbb P}^{(0)}(2 \varepsilon_{\dot{a}\dot{b}}
\overline{\Delta} - \overline{\Delta}_{\dot{a}\dot{b}}); \nonumber \\
Z^{L} & = & {\bar{J}}^{-}_{\dot{1}\dot{2}}\;{\mathbb P}^{o}; \nonumber \\
Z^{R} & = &  {J}^{-}_{ 1 2}\; {\mathbb P}^{o}; \nonumber \\
L_{\mu\nu}^{I} & = &  L_{\mu\nu}^{diag}{\mathbb P}^{e}+  L_{\mu\nu}^{+}{\mathbb 
	P}^{ o}
	\label{eq:gl8/8embeddingI}
\end{eqnarray*}

\section{Conclusions and outlook}
In this paper the ideas of \cite{ModPhysLett} on schizosymmetry in 
spacetime have been developed in detail, and the connection with $sl(2/1, {\mathbb C})_{\mathbb R}$
super-Lorentz symmetry at the component field level has been demonstrated 
for two of the three types of schizofield identified (for the third 
type, the supersymmetry is of orthosymplectic type $osp(3,1/4,{\mathbb 
R})$\cite{Arnowitt}, but was not considered further). 

The question of the 
physical status of such book-keeping supersymmetries remains to be 
established by further extending the formalism to a schizofield 
calculus. A major technical difficulty for the latter is that 
schizofield products are not closed as to type; a bilocal calculus
$$
\Phi \star \Phi'(\theta)  \sim \int_{\vartheta} \Phi(\theta - \vartheta) 
\Phi'(\vartheta) K(\theta, \vartheta) d\vartheta
$$
may be required for covariance, and is under study\footnote{
The embedding of the super-Lorentz algebra in $gl(8/8)$ is 
not regular, leading to non-linear $\theta$ and $\partial^{\theta}$ 
terms in the generators, so that there is no Leibniz
property for handling products of schizofields.}. 
From the super-Lorentz representations established in \S 3, 
namely $\Phi_{I} \simeq ({\mathbf 8},0) \! + \! (0, {\mathbf 8})$, and
$\Phi_{I\!I} \simeq ({\mathbf 4},{\mathbf 4})$, there is a 
natural identification of type $I\!I$ schizofields as `gauge 
potential-like' (containing a vector potential $A_{\mu}$ as one component), and type $I$ 
schizofields as `field strength-like'
(containing an antisymmetric tensor $F_{\mu \nu}$ as one component), 
respectively. A scenario for Lagrangian construction would then be to 
model a 
generalised gauge potential as a Grassmann-odd, type $\widetilde{I\!I}$ schizofield  
$\widetilde{\Phi}_{I\!I}\equiv {\mathcal A}$, say, and to introduce an  
`exterior' operator of the form ${\mathcal D} = \Gamma^{\mu}
\partial^{x}_{\mu}$ for suitable odd $\Gamma^{\mu}$, for example 
$\Gamma^{\mu} = \theta^{\mu}$ or $\Gamma^{\mu} = \gamma^{\mu}$ (see 
(\ref{eq:DiracAlgs})). Local gauge invariance would then be
implemented through
$\mathcal{F} = \mathcal{D} \mathcal{A}  + \mathcal{A} \star \mathcal{A}$ 
(including possible nonabelian extensions)\footnote{
Note for example that $\theta^{\mu}\partial^{x}_{\mu} 
\theta^{\nu}A_{\nu} = \frac 12 \theta^{\mu}\theta^{\nu} F_{\mu \nu}$}.
From this point of view, the present approach can be seen to address 
the question of supersymmetric generalisations of the Dirac operator in 
higher spin wave equations.

\end{document}